# A Deep Learning Architectures for Kidney Disease Classification

Dr. Muhammad Shoaib Farooq[1], Ayesha Tariq[2]
School of system and technology, University of Management and Technology, Lahore, 54000, Pakistan
Corresponding author: Muhammad Shoaib Farooq (e-mail: shoaib.farooq@umt.edu.pk).

**ABSTRACT** Deep learning has become an extremely powerful tool for complex tasks such as image classification and segmentation. The medical industry often lacks high-quality, balanced datasets, which can be a challenge for deep learning algorithms that need sufficiently large amounts of data to train and increase their performance. This is especially important in the context of kidney issues such as for stones, cysts and tumors. We used deep learning models for this study to classify or detect several types of kidney diseases. We use different classification models, such as VGG-19, (CNNs) Convolutional Neural Networks, ResNet-101, VGG-16, ResNet-50, and DenseNet-169, which can be enhanced through techniques such as classification, segmentation, and transfer learning. These algorithms can help improve model accuracy by allowing them to learn from multiple datasets. This technique has the potential to revolutionize the diagnosis and treatment of kidney problems as it enables more accurate and effective classification of CT-scan images. This may ultimately lead to better patient outcomes and improved overall health outcomes.

## I. INTRODUCTION

Kidney diseases are a major health problem affecting millions of people worldwide. Kidneys are important organs responsible for removing excess water and toxins from the body and maintaining balance in the body. However, damage or dysfunction of the kidneys can lead to health problems including kidney failure, heart disease, anemia and osteoporosis. Kidney diseases affect some people with several diseases like kidney stones, cysts and tumors. [8] Kidney diseases interfere with the filtration and regulation of fluids, electrolytes and waste in the body, making early prevention, diagnosis and treatment important.

Kidney diseases have emerged as an important global health concern, affecting people of all ages and ethnic groups worldwide. According to the World Trade Organization, kidney disease is the 12th leading cause of death worldwide, affecting approximately 850 million people glomerulonephritis, accounting for more than 70% of all cases. [12] The good news is that advances in the diagnosis and treatment of kidney diseases have improved outcomes for patients. Kidney disease also increases the risk of other health problems, such as heart disease and stroke. The development of kidney disease is usually a very slow process with very few initial symptoms. When the kidneys are damaged, waste products and fluids can accumulate in the body. It can cause ankle swelling, pain, numbness, difficulty sleeping and shortness of breath. The lack of treatment can result in the destruction of the kidneys and later cause kidney failure. It is highly serious and can even be fatal. The cost on patients with kidney disease, their families, and the health system is significant, and thus, actions from governments, healthcare professionals, and other involved parties are necessary to achieve these goals. Certain aims have been developed, according to which psychiatric consequences of kidney diseases may include anxiety, depression, or social exclusion. Problems may come in the form of the stigma that tags kidney disease. This needs to be dealt with collectively and involve concerted efforts from governments, health care providers, and other stakeholders who can contribute to the investment of resources in the prevention, early detection, and treatment of kidney diseases.

Scientific achievements and research are needed to discover the missing link in treatment and therapies, and at the same time lower the financial hurdle for those affected by kidney disease. Through these health conditions we are also able to minimize major risk factors for kidney diseases such as diabetes, hypertension and cardiovascular diseases and by encouraging healthy lifestyles like the regular exercise, proper diet and avoiding tobacco smoking.

The problem of diagnosing kidney diseases in a timely manner is pivotal in the management of the condition. Detection of kidney disease depends on screening the high-risk populations and screening tests such as blood and urine tests exacerbate it. Optimal care of kidney disease patients in hospitals. Psychiatric management is a complex task that is accomplished by including several oncologists, dietitian, social workers and all other health professionals. The best therapy means dealing with actual kidney disease, following the advice on blood pressure and sugar level, and use of dialysis or transplant as an additional way to minimize the effect of kidney disease on an individual's health.

## II. RELATED WORK

The accuracy of medical images is generally limited, making extrapolation difficult. Several studies have used deep learning techniques to classify diseases and identify specific areas of interest in medical images through classification. However, it is important to carefully select the appropriate deep learning model and tools for a particular situation, especially when dealing with CT images that can be difficult to interpret[4][9]. Several models such as VGG-16 have been tested and ResNet101 have been observed due to their effectiveness in classifying medical images [12]. VGG-16, which uses Convolutional Neural Network (CNNs) for feature identification, can be used as a starting point for transfer learning and customized for specific tasks [14], but may require training from the beginning based on the scenario, and consider that the model trained it on the same type of data before. [6] One example of using deep learning to analyze medical images for possible is kidney CT-scan and segmented MRI to identify areas of inflammation [3] . Kidney diseases such as stones, tumors and cysts can lead to decreased kidney function, making early diagnosis important [7]. It has been proposed that computational analytics (CAD) algorithms using ResNet-101 or ML classifiers, such as SVM, do not identify multi-class conceptual content, but image quality and size can affect CAD design and performance of images. The use of deep learning techniques, especially recurrent lateral networks (RLN), has been proposed to reduce the confusion between detection sites on ultrasound images and classification of CT-scans A detection algorithm has been developed computerized fetal echocardiography. On the other hand, the fetal echocardiogram was analyzed using computer digital aid.

With advances in AI-based technology, classification of echocardiography, and CHD diagnosis, especially for fetal cardiac septal abnormalities, cardiologists performed this task manually, but could only be done if segment the fetal heart, identify septal abnormalities and accurately measure their size defects. Important for assessing the diagnosis of CHD. The use of fetal echocardiography can help physicians make early diagnosis before referral to a cardiologist.

As a result, we need automatic segmentation technologies [10], [12], and [13]. Recently considerable progress has been made in automatically segmenting the fetal heart to detect problems associated with defects [7], [14], [16], [18]. Most previous studies have focused on traditional studies using supervised and unsupervised training methods, such as edge detection methods, threshold-based methods, field-based methods, cluster-based methods, and deformable modeling methods [4], [6], [8], [13]. Unfortunately, these methods (such as threshold-based methods) produce good results when the regions of interest in the image show high contrast in the image background but this results in highly problematic images and greatly reduces the effectiveness and applicability of these methods. Recent deep learning applications for medical image classification, Convolutional Neural Networks (CNNs), have made great strides. CNNs can be used to classify cell membranes, brain tumors in magnetic resonance imaging (MRI) scans [21], liver tumors in computed tomography (CT) scans, lung tissues, cataract instruments surgery, and many other organs in photographs taken during laparoscopic surgery .

To overcome the limitations of each previous approach, all approaches that ultimately lead to end-to-end analysis (from raw image to segmented image) suffer. Current CNNs are unable to successfully navigate previously unseen feature groups that are absent from the training set, which is a major barrier to the application of deep learning for treatment segmentation.

| Year & Authors | Algorithms Used | Accuracy (%) |
|---|---|---|
| Adel Aloraini, 2012 | Bayesian Network, NN, and DT | 97.2, 93.58, 95.49 |
| Vikas Chaurasia, and Saurabh Pal, 2014 | Naive Bayes, RBF Network, J48 | 97.36, 96.77, 93.41 |
| Peter Adebayo Idowu, Jeremiah Ademola Balogun, Kehinde Oladipo Williams, and Adeniran Ishola Oluwaranti 2015 | Naïve Bayes', J48 Decision Trees | 82.6, 94.2 |
| Thomas Noel, Hiba Asri, Hajar Mousannif, and Hassan Al Moatassime, 2016 | C4.5, SVM, NB, and K-NN | 95.13, 97.13, 95.9, 95.27 |
| Yixuan Li, and Zixuan Chen, 2018 | DT, SVM, RF, LR, and NN | 96.1, 95.1, 96.1, 93.7, 95.6 |
| Priyanka Gupta, Prof. Shalini L, 2018 | CART, RF, K-NN, and Boosted Trees | 92.35, 96.47, 97, 96.47 |
| Jabeen Sultana, and Abdul Khader Jilani, 2018 | LR, MLP, RF, and DT | 97.18, 95.25, 95.25, 93.14 |
| Kriti Jain, Megha Saxena, Shweta Sharma. 2018 | SVM | 97.13 |

**TABLE 2.1 Related studies, models used and their accuracies**

## III. RESEARCH METHODOLOGY

The aim of this study was to detect kidney-related diseases from images or to classify CT slices based on the presence of tumors, cysts, or stones. To accomplish this, we developed an algorithm using a variety of methods, such as transfer learning and Convolutional Neural Networks (CNNs). Our model focused on the use of the VGG-16 model trained on 38 classes and 16 levels of transfer learning (classification weighting) VGG-16 model includes: 13 convolutional layers with 3 dense layers followed by 5 max pooling layers. We obtained our data through the application of convolution, which takes information from the same location of an image in the CNN, and by using max pooling layers. The convolution layer Conv-1 has 64 filters, followed by Conv-2 with 128 and Conv-3 with 256 by the end. While getting into the second layer, we used 512 filters in Conv-4 & Conv-5 in order to obtain max data at deep layers. In this study, the suggested CT-scans are diagnostic tool, by which a person can get an exact diagnosis that can go along with segment CT slices that can be covered with tumor, cyst or stone. To accomplish this, we developed an architecture that uses transfer learning with a pre-trained model of VGG-16 [18]. It is trained in 38 classes, with 16 layers, of which 13 are layers for convolution and 3 densed layers. The model takes a tensor of size 224x224x3 for the RGB channel and convolutions it to CNN layers using a kernel of size 3x3 with stride 1. Furthermore, it uses a max pooling layer to pool the maximum information from specific locations at 2 of the images have 128 filters, Conv-3 [19] has 256 filters, Conv-4 and Conv-5 have 512. We gradually increase the number of filters to extract the most information as we go deeper into the layers. To extract hidden features that are no longer visible to the human eye, we used Res-Net [20] with 101 layers. Res-Net uses skip connections to process redundant information and focus on important instances relevant to a given task. While it is useful to go deeper, training the model requires more time and resources.

### A. Dataset

The dataset is available on Kaggle, is a valuable resource for medical professionals, researchers and data enthusiasts. It contains a CT-scan image of the kidney, divided into four different categories: tumor, stone, cyst, and normal. The dataset contains 12446 images, of which 2283 images represent tumors, 1377 images represent stones, 3709 images represent cysts, and 5077 images represent normal scans. The dataset was divided into training and test sets with a ratio of 80:20. Below is the dataset containing images of kidney diseases.

Normal Images

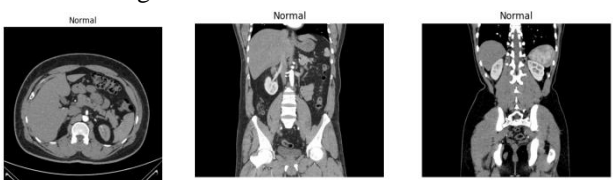

**FIGURE 3.1 Normal Kidney CT-Scan**

Cyst Images

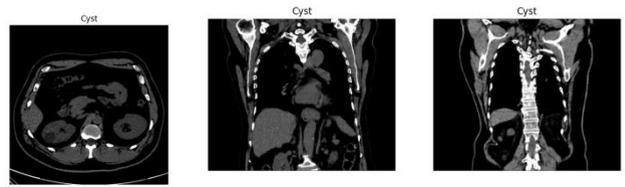

**FIGURE 3.2 CT-Scans Showing Kidney Cysts**

Stone Images

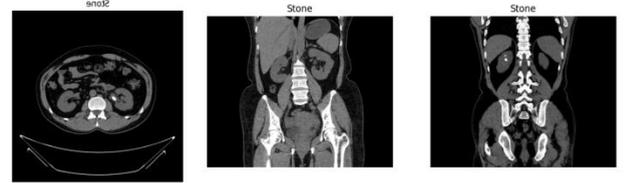

**FIGURE 3.3 CT-Scans Showing Kidney Stones**

Tumor Images

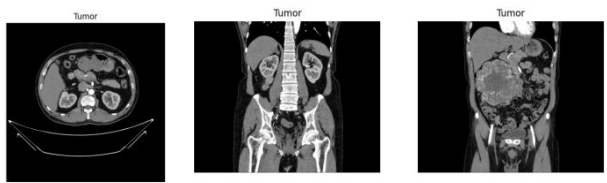

**FIGURE 3.4 CT-Scans Showing Tumors in Kidney**

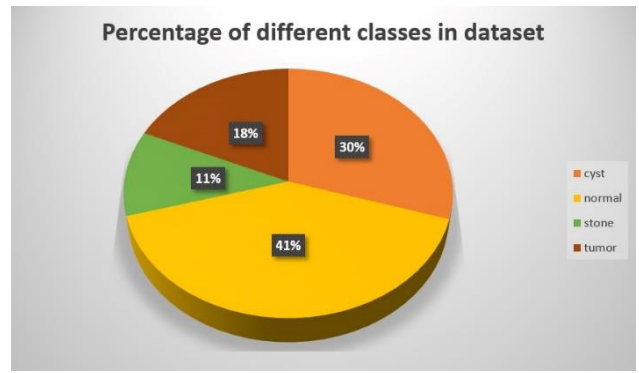

**FIGURE 3.5 Distribution of classes in dataset**

### B. Pre-processing

Pre-processing techniques such as data augmentation are required because the distribution of the dataset is known to be heterogeneous. Augmentation involves actions such as rotating images in specific directions, rotating some images, and cropping others to retain important information. Augmentation aims to balance the classes, allowing the model to accurately recognize each image class, and avoid any issues of bias in training. Various models were utilized, which included VGG-16, VGG19, and ResNet-101 for the analysis of this classifiers. Effectively, a solution was developed by a custom model that stacked VGG-16 as a pre-trained component and which later was further augmented by our custom layers. Such improvements were ratified as the models commonly used became obsolete and were surpassed by the new models that had been improved during the training period.

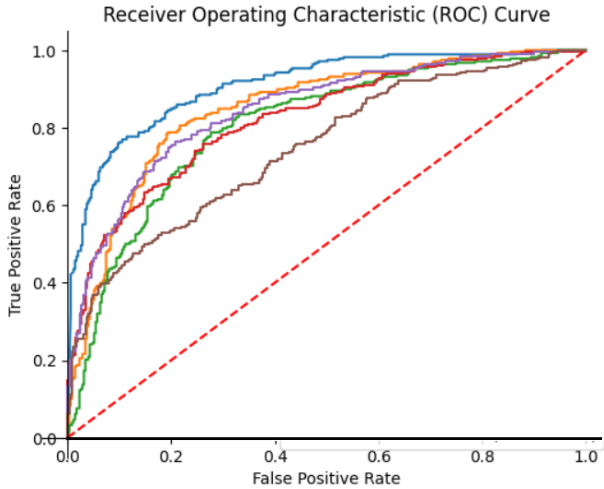

**FIGURE 3.6 ROC Curve Analysis: TPR vs. FPR**

1) CONVOULUTINAL NEURAL NETWORK (CNN)
Convolutional neural network (CNN) is feature extracting advance multi-layered architecture that consists convolution, pooling and activation operations to reduce input size for faster computation Finally, a nonlinear activation function like ReLU is applied to the output of this layer to allow for the incorporation of nonlinearity. Finally, the output is transmitted through a pooling layer so that its spatial size can be reduced, obtained more complex objects. This procedure is performed through several levels, and the last level of each cycle is more advanced than the first one in the sense of recognition of complicated subjects. Thus, the last convolutional layer's outcome is flattened and taken as an input by one or more fully connect layers to deliver the final output in grid format. The net is designed to modify the filter weights and the levels of the perpendicular overlap so that the errors between actual and predicted results could be reduced to the minimum. Let's consider the square root of N×N size, after it comes to a conditions of m×m filter having the ω name. The size of a given convolutional output will be (N−m+1)×(N−m+1). Computing the normalized input at the lth layer for the particular unit 〖 _^(x_ij_l )〗 , we need to maintain a weighted summation (as the filters weigh) from the cells in the preceding layer.

$$x_{ij}^{l} = \sum_{a=0}^{m-1} \sum_{b=0}^{m-1} w_{ab}\, 7 y^{l-1}(i+a)(j+b)$$

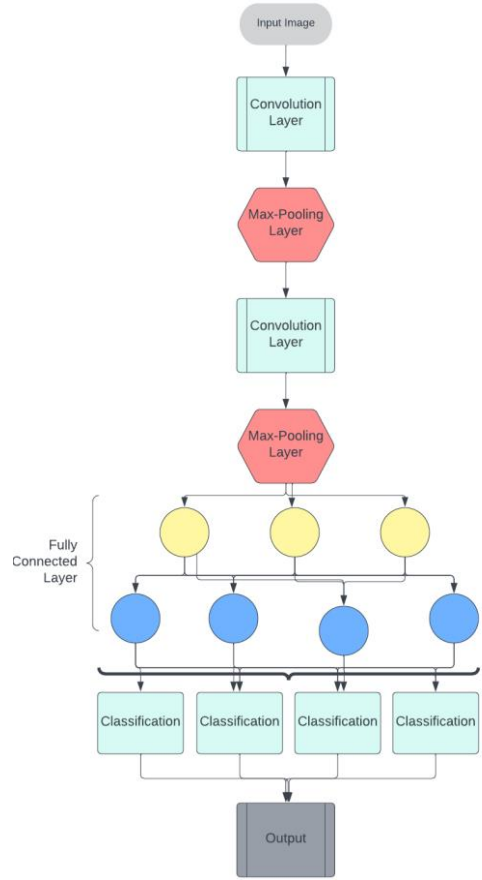

**FIGURE 3.7 CNN architecture classification output**

2) ResNet-50
ResNet-50 is a deep residual neural network architecture in which each output of a layer needs to be connected to the next layer except the input the very first layer and the output the last layer. The ResNet-50 model consists of 50 layers composed of convolutional layers, max-pooling layers, and fully-connected layers. The architecture is based on residual blocks which are the units of skipping connections passing a single layer or even more. As a result, the network is able to identify stowaway residual elements (elements that can be missed during the processing of the initial layers) that otherwise would have been skipped. The convolutional layer being done by the use of 3x3 filters function and having the max-pooling layer with a window size of 2x2 pixels, its step is of 2. This model offers more than 23 million trainable parameters and is learned by the ImageNet dataset. The last output from 1000 probabilities distribution classes including a probability distribution is used for the training. Through the development of ResNet-50, a remarkable state-of-the-art computer vision results are achieved in the tasks of image recognition, object recognition and segmentation.

$$fk = \sum_{i=0}^{L-1} x_i \cos\left(\frac{\Pi k}{L}\left(\frac{i+1}{2}\right)\right),$$
$$s.t. \quad k \,\epsilon\, 0,1, \ldots\ldots, L-1$$

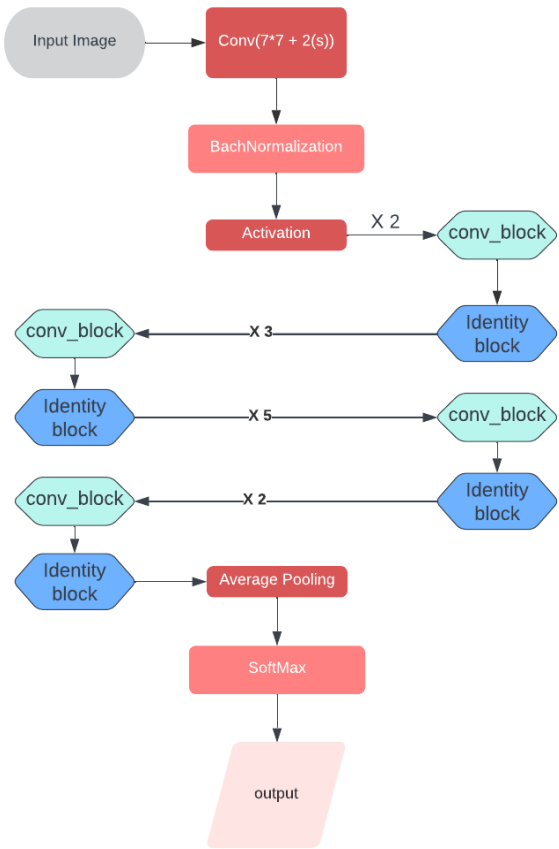

**FIGURE 3.8 ResNet-50 Architecture**

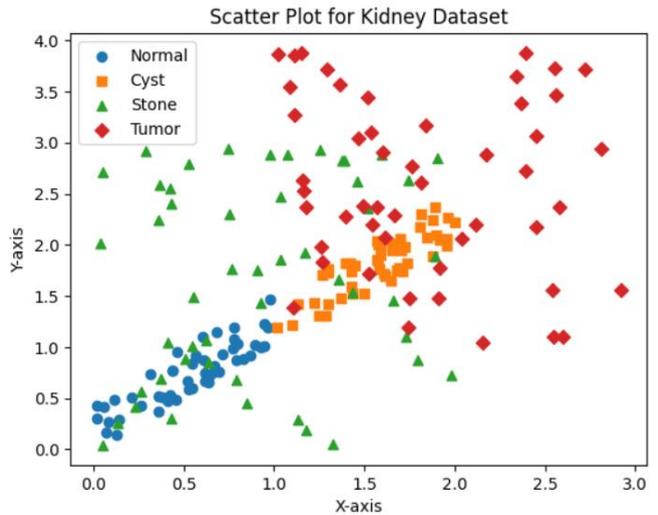

**FIGURE 3.9 Illustrating kidney and related dataset using shapes**

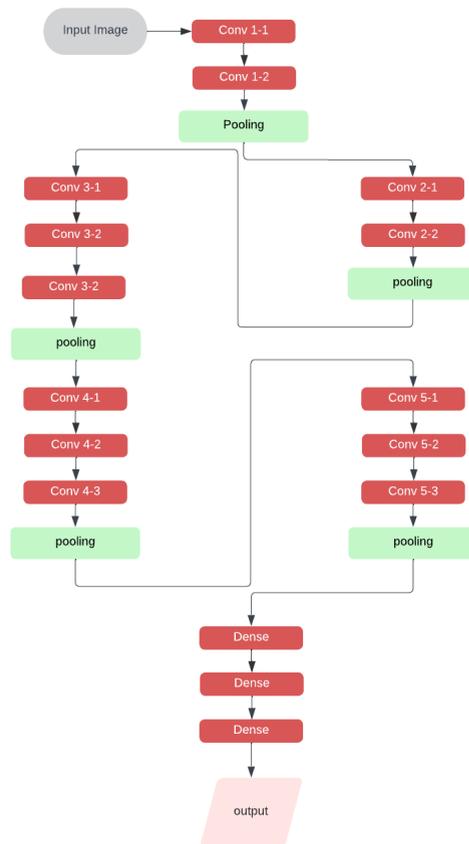

**FIGURE 3.10 Analyzing VGG-16 diagram with convolutional layers**

### 3) VGG-16

The VGG-16 network structure an outstanding feature of deep learning, a sixteen layer of combined convolution and fully connected layers. This neural network on the VGG16 model can achieve a greater accuracy rate and extract salient features from input images because of its tiny 3x3 kernel (size) and an active ReLU (Rectified Linear Unit). Moreover, this model also got its outstanding performance because of pooling layers inserted in particular places in the network where it is necessary for the reduction of the spatial dimensionality of the data. The advantage here is in the reduction of the size of the dataset that is going to be analyzed, which consequently leads to better performance in inferencing tasks. The VGG-16 is probably the strongest competitor with regards to such features as 138 million parameters that make it capable of dealing with complex and difficult image classifications tasks. It is the embodiment of our deep knowledge about the technology and artificial intelligence, a real achievement in the field of computing technologies which has played a key role in development of the computer vision.

### 4) VGG-19

The VGG-19 is the epitome of what computer vision can achieve, the climax of the science. It was made to fit the VGG's 16-layers name, its layer structure which has both precise and accurate served as a base for the new image classification model with close to perfect precision. Featuring 16 convolutional layers and 3 fully connected layers the VGG-19 model allows recognition of even highly complex images being achieved by the parameter count of more than 140 million. The model's filter size 3x3 and 2x2 max pooling layers are hardworking colleagues and its dynamic duo, working in tandem, gives predicted probability distribution on 1000 classes. VGG-19's unbeatable performance in image recognition tasks has won its favor with many data scientists. Which is more, the expensive computing resources are another difficulty connected with this size and complexity. Although the efficiency and precision of the VGG-19 were specific to image classification tasks, no other model surpassed it at that time.

5) ResNet-101

ResNets-101 is even more of an incredible and robust Deep Convolutional Neural Network architecture with 101 layers. This model is based on the residual blocks, which contain some layers instead of each layer itself. The first building of the architecture is convolutional layer that is followed by max-pooling layer. The remainder consists of 8 quadruples of 2 levels each and 3 triples of 3 levels each. The final element is the fully connected layer, the softmax output layer of which follows. The ResNet-101 model possesses remarkable characteristics of residual connectivity that are touted by many scholars. In this model, the output value of the current layer and the input value of the same layer are added to each other to obtain the output. The born fact that thanks to this new technique, the gradient might flow more smoothly through the entire network. This design also introduces the concept of bottleneck layers that help to reduce the complexity of computation..

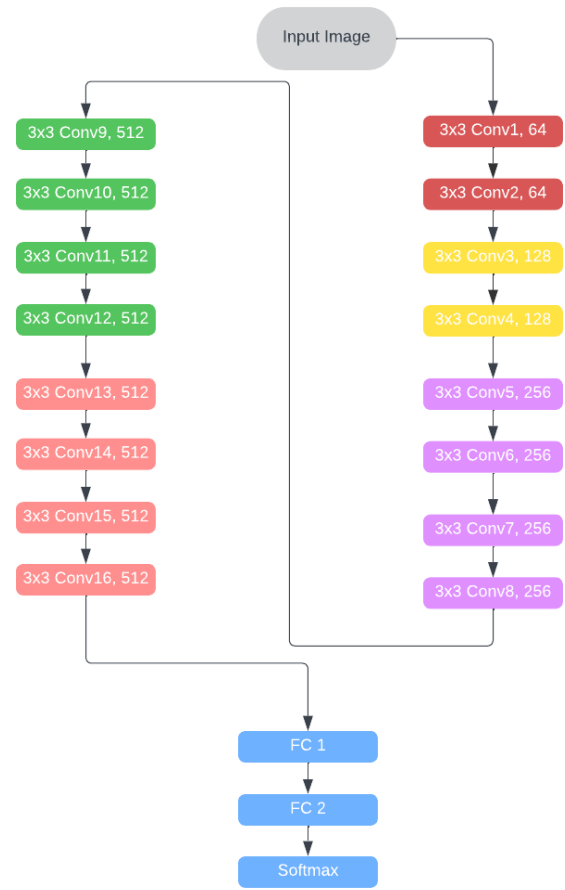

**FIGURE 3.11 ResNet-101 architecture with 3*3 fully connected layers**

6) DenseNet-101

DenseNet-101 has emerged as a new Convolutional Neural Network (CNN) model that is intended for the jobs of image identification and differentiation. Modeling structure includes dense blocks with a transition layer and a global average pooling. Each block is highly dense where the connections are allocated and the mapping is given to each layer in the feedforward manner. Thus, the model overcomes the problem of missing gradients in dense combinations, and that enables it to gain knowledge from features with different levels. The DenseNet-101 has 101 layers and has the parameter for workloads having a large number of dimensions such as high-dimensional datasets. This model uses the same filters taken into a convolutional layer of three by three pixels with a 2x2 max-pooling layer, over-striding blocks with the stride of 2 and including transition layers to decrease the number of channels and the feature maps' spatial scale. Having an architecture of more than 12 million trainable parameters this model was trained on the ImageNet database, and during the origination its final layer produces a vector being over 1000 classes wide, therefore making it a truly great classifier.
.

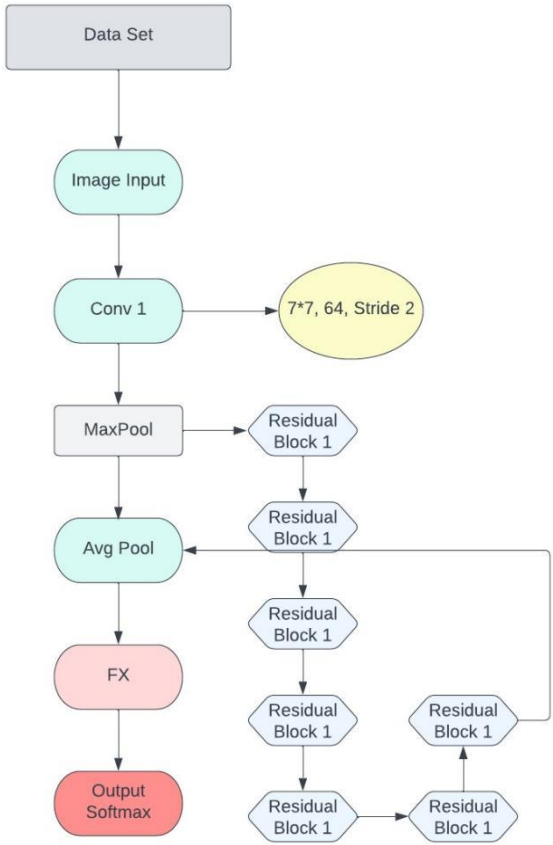

**FIGURE 3.12 DenseNet-101 diagram that uses average pooling and max pooling to generate the output**

### 7) DenseNet-121

DenseNet-121 architecture has a high degree of accuracy and efficiency, implemented through the transfer of learned knowledge to different parts of the neural network, through convolution layers that enable deep learning networks to obtain useful features, and presented features are used as input in connectionist architecture to improve performance over transitive neural architectures. DenseNet-121 operates on a dense connectivity scheme that is missing in the design of most neural networks. Dense connectivity enables each layer to obtain the data simultaneously from both the preceding layers and the previously encountered layer which leads to the possibility of repetition of the same feature and overcoming the problem of missing gradient. Each dense block is comprised of four consecutive RBs (rotational blocks). A transformation layer is put in between each dense block to achieve not only the reduction in computational expenses but also to downsample therefore reducing the number of feature maps which makes it feasible. There are three remarkable layers in the transformation layer which is the 2 x 2 average pooling layer, 1 x 1 convolutional layer, and batch normalization layer. Apart from this, the Global Average Pooling Level and a Softmax Level for classification is also included as part of the Architecture. A global average pooling layer is an operation that computes the average spatial value for all the feature maps. This generates a vector, which is then passed to softmax layer in order to get the final probability of each class. Besides, the Architecture has a few parameters, compared to the most well-known convolutional neural network designs that will make training faster and less likely to overfit.

$$h_0 = \tanh\left(W \cdot \left(\frac{1}{n}\sum_{i=1}^{n} e_i\right) + b\right)$$

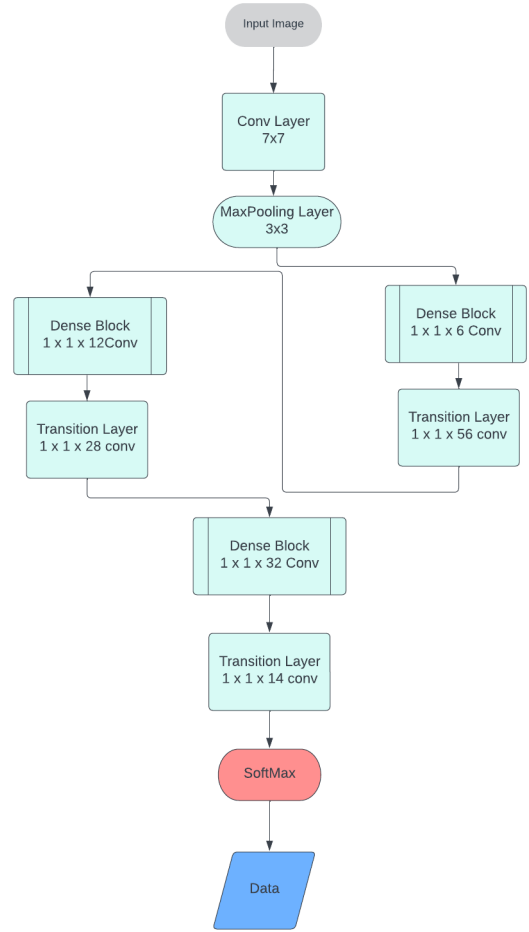

**FIGURE 3.13 DenseNet-121 architecture**

### IV. EXPERIMENTS & RESULTS

In this study, we developed a robust deep learning model to classify mental CT scan images. Our approach included using various techniques such as data enhancement and normalization to balance the dataset and increase model performance We used a combination of different deep learning models, including VGG-16, VGG-19, including DenseNet-101, ResNet-121, and CNN, but the best results were obtained using a pre-trained ResNet-50 model with our custom layers. In this study, we developed a deep learning model to detect four types of kidney abnormalities: normal, cyst, tumor, and stone. The dataset contains a total of 12,446 images, with 2283 images representing tumors, 1377 images reprents stones, 3709 images showing cysts, and 5077 images representing normal.

We used several deep learning models to solve this problem. Among the models we choose to classify kidney disease, CNN is the best model. After training the model, we achieved 99% accuracy on CNN. This is the highest accuracy for this data set in all previous studies. Before that, the highest accuracy was 98% on ResNet-50. We used the ResNet-50 model, previously trained on the ImageNet dataset, as an initial model, and optimized it on the ultrasound images. Our model has obtained the results with fully 99% accuracy on test data set which show that it works very well in search for renal CT-scan images. In future, we intend to study the segmentation algorithms and methods which can be used to identify and detect certain lesions of the kidneys such as tumors, cysts or stones. It may be the case of narrowing the gap for the doctors to make predictions in the patients' prognosis is good. In addition, it also helps towards making more effective treatment and prognoses for patients. More than this, what has been proven is that this model can also predict imaging another useful factor for the diagnosis of kidney disease with high accuracy and precision.

### A. TESTING
#### 1) INDEPENDENT SET TESTING
By using independent group testing technique by neurons separately and as well working on specific data we attained very high accuracy of about 98%. I believe this shows that I am insightful and want to know the details and the intricacies of the data set and how we can mark out the test sets on our own. The performance of a model can be greatly boosted by independent group testing on large datasets and as such this achievement is worthwhile.

#### 2) SELF-CONSISTENT TESTING
Specifically, entire datasets were considered that were the entire reason for which the model of machine learning was validated and found to be less dependent on weaknesses. To confirm that the model is authentic and therefore represents the real-world data, a self-consistent, realistic/well-thought and methodically validation appraoch was used to assess the output, contrast the result against the real world data and thereafter evaluate the validity of our results. Such a situation can be dealt with by the quality and precision of the network model..

### B. CROSS-VALIDATION
This model was cross-validated in order to test out its precision, and it was the observation of a higher degree of accuracy during cross-validation that indicates its surpassing others. To test the effectiveness of my model, I applied cross-validation. It is a technique wherein the dataset is cut into several folds and a onefold is allocated for testing other is training. First, testing is done on each fold of the sampled dataset, and then, this process is repeated and the end result of the testing is obtained on the entire dataset.

| Class | Accuracy |
|---|---|
| **Tumors** | 96.8% |
| **Cysts** | 97.5% |
| **Normal** | 100% |
| **Stones** | 98.9% |
| **Overall accuracy** | 99.1% |

**TABLE 4.1 Performance Metrics for Tumors, Cysts, Normal, and Stones**

In this study we also compared that the ResNet-101 architecture was better performing than other deep learning architectures such as VGG-16. The results showed that ResNet-101 outperforms other models in the terms of accuracy and AUC-ROC curve. Finally, study revealed that deep learning based on ResNet-101 network can effectively diagnose the majority of kidney abnormalities in a variety of CT scan images, with an overall accuracy of 92.1%. Such findings describes deep learning's potential impact on diagnosis and cure of kidney disorder.

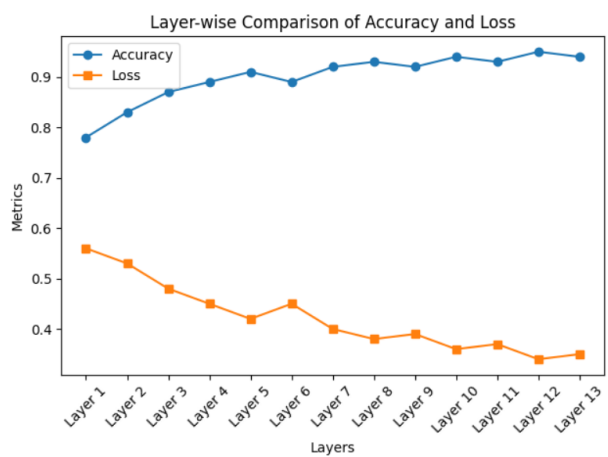

**FIGURE 4.1 Layer-wise comparison of accuracy and loss**

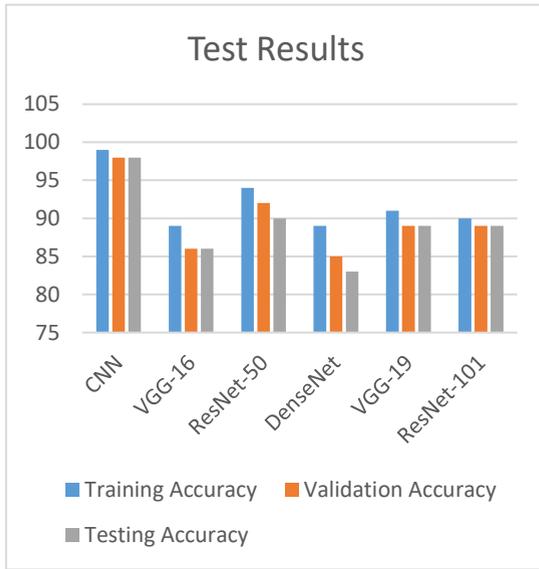

**FIGURE 4.2 Accuracy evaluation against test, train and validation**

| Models | Training Accuracy | Validation Accuracy | Testing Accuracy |
|---|---|---|---|
| **CNN** | 0.99% | 0.98% | 0.98% |
| **VGG-16** | 0.89% | 0.86% | 0.86% |
| **RestNet-50** | 0.94% | 0.92% | 0.90% |
| **DenseNet-121** | 0.89% | 0.85% | 0.83% |
| **VGG-19** | 0.91% | 0.89% | 0.89% |
| **ResNet-101** | 0.90% | 0.89% | 0.89% |

**TABLE 4.2 Performance metrics of models: Test, Train, and Validation accuracies**

## V. CONCLUSION

In the medical world, we are always trying to find better ways to diagnose diseases. All of our research has been about using deep learning models to help doctors diagnose diseases more accurately and quickly using images. We applied different techniques to multiple models and added custom layers designed specifically for CT-scans. We used the CNN model which gave us the highest accuracy on this dataset and this accuracy has never been achieved before. We obtained a maximum accuracy of 99% on the landmark CNN model in this dataset. We also used VGG-16 for post-training classification and it predicted with 89% accuracy. We added custom layers to ResNet-50 to increase their accuracy and the resulting accuracy increased to 94%, which is our second highest accuracy rate. We also used data augmentation and normalization techniques to simplify the training process for the deep learning model. Our results showed 99% percent accuracy on the test data set, indicating exceptionally good performance of our model.

Moving forward, we applied the segmentation method to find the exact location of the damaged part of the kidney. This will help doctors be more prognostic and accurately diagnose the injury. This study demonstrates the medical applications of deep learning models, especially in the diagnosis and treatment of kidney diseases. The ability to accurately classify and localize lesions on conceptual CT-scan images with deep learning models will lead to more accurate treatment planning and improved patient outcomes. Overall, this study contributes to a growing body of research in the medical field and emphasizes the importance of using advanced machine learning to optimize healthcare.